\documentclass[12]{iopart}

\usepackage{amssymb}
\usepackage{epsfig}

 \expandafter\let\csname equation*\endcsname\relax
 \expandafter\let\csname endequation*\endcsname\relax 
\usepackage{amsmath}
\usepackage{bbm}
\usepackage{float}

\newtheorem{defi}{Definition}

\newcommand {\be}{\begin{equation}}
\newcommand {\ee}{\end{equation}}
\newcommand {\bey}{\begin{eqnarray}}
\newcommand {\eey}{\end{eqnarray}}

\newcommand{\noi}{\noindent}
\newcommand{\cancel}[1]{}
\newcommand{\canpro}[1]{}

\newcommand{\bea}{\begin{eqnarray}}
\newcommand{\beq}{\begin{equation}}
\newcommand{\eea}{\end{eqnarray}}
\newcommand{\beas}{\begin{eqnarray*}}
\newcommand{\eeas}{\end{eqnarray*}}
\newcommand{\bece}{\begin{center}}
\newcommand{\ence}{\end{center}}
\newcommand{\beit}{\begin{itemize}}
\newcommand{\enit}{\end{itemize}}

 \def\<{\langle}
 \def\>{\rangle}

 \def\opone{\leavevmode\hbox{\small1\kern-3.8pt\normalsize1}}

 \newcommand{\complex}{{\kern .1em {\raise .47ex\hbox {$\scriptscriptstyle
 |$}}\kern -.4em {\rm C}}}
 \newcommand{\real}{{{\rm I} \kern -.19em {\rm R}}}
\newcommand{\eeq}{\end{equation}}
 \newcommand{\beqa}{\begin{eqnarray}}
 \newcommand{\eeqa}{\end{eqnarray}}

\begin{document}

\title{Non-Locality of Experimental Qutrit Pairs}

%\author[1]{C. Bernhard}
%\author[1]{B. Bessire}
%\author[2]{A. Montina}
%\author[2]{M. Pfaffhauser}
%\author[1]{A. Stefanov}
%\author[2]{S. Wolf}
%\affil[1]{Institute of Applied Physics, University of Bern, Bern, Switzerland.}
%\affil[2]{Department of Informatics, USI Lugano, Switzerland.}

\author{C~Bernhard$^{\dagger}$, B~Bessire$^{\dagger}$, A~Montina$^{*}$, M~Pfaffhauser$^{*}$, A~Stefanov$^{\dagger}$, S~Wolf$^{*}$}
\address{$^{\dagger}$ Institute of Applied Physics, University of Bern, Bern, Switzerland.} 
\address{$^{*}$ Faculty of Informatics, USI Lugano, Switzerland.}

%\author{C.~Bernhard\S, B.~Bessire\S, A.~Montina\P, M.~Pfaffhauser\P, A.~Stefanov\S, S.~Wolf\P}
% \address{\S Institute of Applied Physics, University of Bern, Bern, Switzerland.} 
% \address{\P Faculty of Informatics, USI Lugano, Switzerland.} 
%\date{}

%\maketitle

%\vspace*{-0.7cm}

\begin{abstract}
\noi
The insight due to John Bell that the joint behavior of individually measured entangled quantum systems cannot be explained by shared information
remains a mystery to this day. We describe an experiment, and its analysis,
displaying non-locality of entangled qutrit pairs. The non-locality of such
systems, as compared to qubit pairs, is of particular interest since it potentially opens the door for tests
of bipartite non-local behavior independent of probabilistic Bell inequalities,
but of deterministic nature.
\end{abstract}

\section{Non-Locality in Theory \ldots}
%Author: Stefan
%\begin{itemize}
%\item A long way since Bell. (Main theme a lot happened in the last 50 years).
%\item New Bell inequalities
%\item qubit $\to$ Qudit
%\item Theory $\to$ also Experiments
%\item Bell violations $\to$ new methods
%\item sequence: CHSH $\to$ Newer Bell Inequalities $\to$ Higher Dimension $\to$ also experiments $\to$ other methods to show non-locality
%\end{itemize}

The discovery of {\em John Bell\/} that nature was non-local and quantum physics cannot
be embedded into a local realistic theory in which the outcomes of experiments
are completely determined by values located where the experiment actually takes place,
and no far-away variables, is profound. It
probably poses more questions than it has solved: If shared classical information~\cite{Bell1964}
as well as hidden communication~\cite{transitiv} both fall short of explaining non-local correlations,
what ``mechanism'' could possibly be behind this strange effect? What can we learn
from Bell's insight | and from the series of experiments carried out in
the aftermath~\cite{Freedman1972,Aspect1982a,Tittel1998,Weihs1998} |
about the nature of space and time? What can we conclude about free choices (randomness)
and, more generally, the role of {\em information\/} in physics and in our understanding of
natural laws? In fact, Bell's theorem is an {\em information-theoretic\/} statement:
It characterizes what kind of joint input-output behavior
\[
P(ab|xy)
\]
can in principle be explained by shared (classical) information $R$, i.e., is of the form
\[
\sum_r{P_R(r) P^r(a|x)P^r(b|y)}\ ,
\]
and which cannot.
The relationship to (quantum) physics is given by the fact that the joint behavior ---
where the inputs are given by
the choice of the measurement settings and the output is the measurement outcome --- of  two (or more)
parts of some entangled system are non-local, i.e., violate a {\em Bell inequality}.
Bell inequalities are linear constraints that all local systems satisfy and that, in
their totality, define the {\em local polytope\/} in the space of {\em non-signaling
systems}, the latter being all behaviors not allowing for the transmission of information.

In this article, an experiment is described in which entangled three-dimensional systems
are generated and measured, and it is analyzed whether non-local behavior can be observed.
More precisely, the experiment, and its analysis, deals with maximally as well as
partially entangled qutrit pairs. One of the questions of interest is whether maximal
entanglement also means maximal non-locality. Hereby, the strength of non-locality
is measured not only by the extent of a Bell inequality violation, but also in terms
of the distance to the local polytope, as well
as a novel method based on the amount of  {\em communication\/}
required for classically simulating the behavior observed experimentally.

The experimental realization of entangled qutrit pairs as opposed  to qubit
pairs is of interest for several reasons. One of them is that in principle, non-local
correlations directly based on impossible colorings, i.e., the Kochen/Specker theorem~\cite{KS},
become possible~\cite{RenWol}. This type of non-local behavior is conceptually simpler since it does
not depend on probabilities, but is deterministic in nature.

%\section{Preliminaries}

\section{\ldots\ and Experiment}
%Author: Berne
%\begin{itemize}
%\item Starting with a Bell inequality since it all started with them.
%\item Stating the inequality
%\item Stating the properties (maximally violated for non symmetric state).
%\end{itemize}

Due to the extremely weak interaction with their environment, photons exhibit strong resistance against decoherence and can, therefore, be transmitted over large distances without altering their state. Photonic quantum states are thus ideal systems to carry out tests of non-locality with entangled states. Moreover, photons entangled in their polarization, momentum, or energy-time degrees of freedom can be experimentally generated by non-linear processes and coherently controlled and modified through linear optics.
In the context of non-locality tests, the experimental study of entangled photonic states started in 1972 with the observation of non-classical correlations in the polarization of two photons emitted in an atomic cascade from a calcium atomic effusive beam \cite{Freedman1972}. This type of source was further improved and an experiment was performed in 1982 where the changes of the analyzers' orientation and the detection on each side were separated by a spacelike interval \cite{Aspect1982a}. This experiment showed a statistically significant violation of Bell's inequality, demonstrating the non-locality of nature at its fundamental level. Since then, non-locality has been tested by numerous experiments which all confirmed the predictions of quantum mechanics under different conditions. Theories reconciliating the quantum collapse and Lorentz invariance have been tested \cite{Stefanov2002}. Further, boundaries on the speed of a potential superluminal influence have been experimentally set \cite{Salart2008}. Entanglement between photons was shown to be conserved over distances on the order of kilometers in optical fibers \cite{Tittel1998} or in free-space \cite{Ursin2007}. Ongoing research is oriented towards experiments between ground station and low-orbit satellites.

Much effort is nowadays put into the closure of possible loopholes
which can distort the outcomes of Bell test measurements. The presence
of loopholes allows to explain the measurement results by a local
theory even if a measured violation of a Bell inequality pretends the
existence of non-classical correlations. 
Apart from space-time separation between the measurement devices (the
communication loophole), 
the most studied loophole is the so called detection loophole. 
This loophole can be closed using detectors with sufficiently high efficiencies. Massive particles like ions or atoms can be detected with an efficiency of almost unity \cite{Rowe2001a}, however, have not yet been detected in a spacelike separated configuration \cite{Hofmann2012}. Photons are nowadays detected with high efficiency in Bell experiments using super-conducting single photon detectors \cite{Giustina2013}.

The locality condition is very difficult to be fully satisfied without
any additional hypothesis on the theory to be tested. It requires that
the choices of the measurement settings have to be realized without any
possible local causal relation between them. The most independent
choices would be that Alice and Bob freely chose the settings
themselves. The same applies for the detection of the  measurement results. Because the measurement problem is still not solved in quantum mechanics, the detection should be considered to occur at the instant when Alice and Bob have conscience of the outcome result. In both cases the reaction times of Alice and Bob would require to perform the experiment over distances larger than the earth's diameter.
   
Most of the experiments mentioned here were carried out with photon
pairs in entangled qubit states. Their non-locality was revealed by
violating a Bell inequality, for instance the CHSH 
inequality~\cite{Clauser1969}, given by 
\begin{equation}\label{eq:CHSH}
I_{2}\le2\ .
\end{equation}
In the following, we let $A_x$, with $x=1,2$, be an operator measured by Alice and~$B_y$, with $y=1,2$, be an operator measured by Bob. The Bell parameter $I_2$ is then expressed as
\begin{equation}
\begin{split}
I_{2}=|P\left(A_{1}=B_{1}\right)-P\left(A_{1}\neq B_{1}\right)+P\left(A_{1}=B_{2}\right)-P\left(A_{1}\neq B_{2}\right) \\ 
-P\left(A_{2}=B_{1}\right) +P\left(A_{2}\neq
  B_{1}\right)+P\left(A_{2}=B_{2}\right)-P\left(A_{2}\neq
  B_{2}\right)|\ ,
\end{split}
\end{equation}
and the outputs $a$ and $b$ are restricted to $a,b=0,1$. The maximal violation of Eq.~\eqref{eq:CHSH}, together with the corresponding measurement settings, can be fully determined by means of the two-qubit density matrix and the algorithm provided in~\cite{Horodecki1995}. The maximal violation of the CHSH inequality is achieved for a maximally entangled state given by
\begin{equation}
\left|\psi\right\rangle^{(2)}
=\frac{1}{\sqrt{2}}(\left|0\right\rangle_A\left|0\right\rangle_B
+\left|1\right\rangle_A\left|1\right\rangle_B)\ .
\end{equation}

However, nothing restrains us from using pairs of photons entangled in
more-than-two-dimensional spaces. While the polarization of light only
carries two orthogonal states, the other modes of light, energy, and transverse momentum, can in principle encode much larger dimensions. Whereas the behavior of entangled qubits is fully characterized, their extensions to higher dimensions, denoted as qudits, are far from having been fully studied. A first approach to examine the non-classical correlations between two entangled qudits is to consider these states in the context of a new family of Bell inequalities introduced by Collins \textit{et al.}~(hereafter referred to as CGLMP) in~\cite{Collins2002}.

%\section{Setup}

\subsection{Bell inequality for entangled qutrits}
%Author: Berne
%\begin{itemize}
%\item Description of the states
%\item Description of the measurement bases
%\item Why this bases and states?
%\end{itemize}

In the following, we focus on a two-qutrit state
\begin{equation}\label{eq:qutrit_gamma}
\left|\psi\right\rangle^{(3)}(\gamma) =\frac{1}{\sqrt{2+\gamma^{2}}}\left(\left|0\right\rangle_A\left|0\right\rangle_B +\gamma\left|1\right\rangle_A\left|1\right\rangle_B +\left|2\right\rangle_A\left|2\right\rangle_B \right)
\end{equation}
with a variable degree of entanglement $\gamma$.   

In order to experimentally investigate the non-local properties of the  state~(\ref{eq:qutrit_gamma}) we make use of the CGLMP inequality for $d=3$ given by
\begin{equation}\label{eq:CGLMP}
I_{3}\le2
\end{equation}
with the corresponding Bell parameter
\begin{equation}\label{eq:I3}
\begin{split}
I_{3}=\vert P\left(A_{1}=B_{1}\right)+P\left(B_{1}=A_{2}+1\right)+P\left(A_{2}=B_{2}\right)+P\left(B_{2}=A_{1}\right) \\
-P\left(A_{1}=B_{1}-1\right)-P\left(B_{1}=A_{2}\right)-P\left(A_{2}=B_{2}-1\right)-P\left(B_{2}=A_{1}-1\right)\vert\
.
\end{split}
\end{equation}
For $d=3$, we have measurement outcomes $a,b=0,1,2$. At first sight, it seems to be intuitive that a maximally entangled state leads to the highest violation of a Bell inequality. However, for entangled qudits with $d>2$, there exists theoretical evidence that non-maximally entangled qudits reach higher violations of the CGLMP inequality than maximally entangled states \cite{Acin2002,Zohren2008}. In fact, it was shown that Eq.~\eqref{eq:CGLMP} is maximally violated for $\gamma=\gamma_{max}\thickapprox0.792$. To experimentally determine $I_3(\gamma)$, we exploit energy-time entangled photons to encode entangled two-qutrit states according to Eq.~\eqref{eq:qutrit_gamma}. 
%The measurement settings
%use to maximally violate the inequality are the basis
%\begin{eqnarray*}
%A_{1}:\left\{ \frac{\left|0\right\rangle +\left|1\right\rangle +\left|2\right\rangle }{\sqrt{3}},\frac{\left|0\right\rangle +e^{i\frac{2\pi}{3}}\left|1\right\rangle +e^{i\frac{4\pi}{3}}\left|2\right\rangle }{\sqrt{3}},\frac{\left|0\right\rangle +e^{i\frac{4\pi}{3}}\left|1\right\rangle +e^{i\frac{2\pi}{3}}\left|2\right\rangle }{\sqrt{3}}\right\} \\
%A_{2}:\left\{ \frac{\left|0\right\rangle +e^{i\frac{\pi}{3}}\left|1\right\rangle +e^{i\frac{2\pi}{3}}\left|2\right\rangle }{\sqrt{3}},\frac{\left|0\right\rangle -\left|1\right\rangle +\left|2\right\rangle }{\sqrt{3}},\frac{\left|0\right\rangle +e^{i\frac{5\pi}{3}}\left|1\right\rangle +e^{i\frac{10\pi}{3}}\left|2\right\rangle }{\sqrt{3}}\right\} \\
%B_{1}:\left\{ \frac{\left|0\right\rangle +e^{i\frac{\pi}{6}}\left|1\right\rangle +e^{i\frac{\pi}{3}}\left|2\right\rangle }{\sqrt{3}},\frac{\left|0\right\rangle -i\left|1\right\rangle -\left|2\right\rangle }{\sqrt{3}},\frac{\left|0\right\rangle +e^{-i\frac{7\pi}{6}}\left|1\right\rangle +e^{-i\frac{\pi}{3}}\left|2\right\rangle }{\sqrt{3}}\right\} \\
%B_{2}:\left\{ \frac{\left|0\right\rangle +e^{-i\frac{\pi}{6}}\left|1\right\rangle +e^{-i\frac{\pi}{3}}\left|2\right\rangle }{\sqrt{3}},\frac{\left|0\right\rangle -e^{-i\frac{5\pi}{6}}\left|1\right\rangle -e^{-i\frac{5\pi}{3}}\left|2\right\rangle }{\sqrt{3}},\frac{\left|0\right\rangle -i\left|1\right\rangle -\left|2\right\rangle }{\sqrt{3}}\right\} 
%\end{eqnarray*}

\subsection{Experiment}
We use an experimental setup in which energy-time entangled photons are generated by means of a spontaneous parametric down-conversion (SPDC) process induced by a quasi-monochromatic pump laser. It can be shown with numerical methods \cite{Wihler2012} that in this case, the entanglement content in the generated two-photon state is potentially very high by calculating an entropy of entanglement of $E=(21.1\pm 0.2)$ ebits.

\begin{figure}[hbt]
\centering \fbox{\epsfig{file=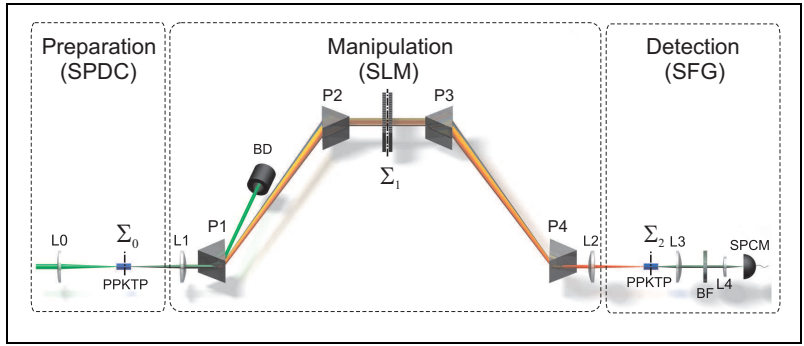,width=0.79\textwidth}}
\caption{Overview of the experimental setup. Preparation: L0 pump beam lens with focal length $f$ = 150 mm, PPKTP non-linear crystal for SPDC. Manipulation: BD beam dump, SLM spatial light modulator, symmetric two-lens (L1, L2) imaging arrangement ($f$ = 100 mm) with a (de)magnification factor of six, P1-P4 four-prism compressor. Detection: PPKTP non-linear crystal for SFG, BF bandpass filter, SPCM single photon counting module with a two-lens (L3, L4) imaging system. Imaging planes are denoted with $\Sigma_k$. \label{fig:overview_setup}}
\end{figure}

The experimental arrangement shown in Figure \ref{fig:overview_setup} is subdivided into three parts: The SPDC crystal prepares entangled two-photon states whose spectrum is discretized with a spatial light modulator (SLM) to generate entangled qudits. The ability of the SLM to individually control the amplitude and phase of selected spectral components then allows to manipulate the qudit states. Finally, the photons are detected in coincidence through sum-frequency generation (SFG) in a second non-linear crystal. 

To prepare entangled photon pairs degenerated at 1064 nm, we pump a $L_{DC}=11.5$ mm long and periodically poled KTiOPO$_{4}$ (PPTKP) crystal with a poling periodicity of $G_{DC}=9$ $\mu$m. The pump laser is a quasi-monochromatic Nd:YVO$_{4}$ (Verdi) laser centered at 532 nm featuring a narrow spectral full width half maximum of about 5~MHz. The collinear pump beam is focused into the middle of the PPKTP crystal ($\Sigma_0$) with a power of 5 W. The down-conversion crystal is mounted in a copper block whose temperature is stabilized to $\pm 0.1$~$^\circ$C. The operating temperature of the PPKTP crystal is chosen for collinear emission and, according to type-0 SPDC, all involved photons are identically polarized. This configuration leads to a spectral width of $\Delta \lambda_{DC}\approx 105$~nm around 1064 nm which implies a spectral mode density of $n\approx 0.2$ entangled pairs per mode \cite{dayan_silberberg05prl}. Therefore, we operate below the single photon limit, and thus in the quantum regime, where the entangled pairs are temporally well separated from each other. The corresponding two-photon state can be derived by first-order perturbation theory and, under the assumption of a quasi-monochromatic pump field, is described by 
\begin{equation}\label{eq:DC_state}
\vert\psi\rangle = \int_{-\infty}^{\infty}d\omega
\Lambda(\omega)\hat{a}^{\dagger}_{i}(\omega)\hat{a}^{\dagger}_{s}(-\omega)\vert0\rangle_i\vert0\rangle_s\
,
\end{equation}
where we omit the leading order vacuum state. Signal and idler photons are created with corresponding relative frequency $\omega$ by acting with the operators $\hat{a}_{i,s}^{\dagger}(\omega)$ on the composite vacuum state $\vert0\rangle_i\vert0\rangle_s$. The joint spectral amplitude reads
\begin{equation}\label{eq:lambda}
\Lambda(\omega)\propto  \mathrm{sinc}\left[\frac{\left(\Delta k_{DC}(\omega)+\frac{2\pi}{G_{DC}}\right)L_{DC}}{2}\right],
\end{equation}
where the phase mismatch $\Delta k_{DC}(\omega)$ is responsible for the efficiency of the SPDC process and includes the dispersion properties of the down-conversion crystal through its corresponding Sellmeier equations.

To manipulate their spectrum, the entangled photons are imaged from the SPDC crystal through a four-prism compressor (P1-P4) to the plane $\Sigma_{2}$. The prism compressor consists of four equilateral N-SF 11 prisms arranged in minimum deviation geometry. The lens L1 images the plane $\Sigma_{0}$ to $\Sigma_{1}$ such that the spectral components are spatially dispersed in order to form a quasi-Fourier plane. Placing a spatial light modulator (SLM, Jenoptik SLM-S640d) at $\Sigma_{1}$ then allows to individually manipulate the spectral amplitude and phase of the entangled photons similar to pulse shaping techniques applied to classical femtosecond laser pulses \cite{Weiner2000}. The effect of the SLM on each photon is described by a complex transfer function $M^{i,s}(\omega)$ which transforms the joint spectral amplitude of Eq.~\eqref{eq:lambda} according to
\begin{equation}
\tilde{\Lambda}(\omega)=M(\omega)\Lambda(\omega)
\end{equation}
with
\begin{equation}
M(\omega)=M^i(\omega)M^s(-\omega)\ .
\end{equation}

Instead of using two spatially separated single-photon detectors, coincidences of the entangled photon pairs are measured by SFG in a second PPKTP crystal with length $L_{SFG}=11.5$ mm and poling period $G_{SFG}=9$ $\mu$m. This ensures an ultrafast coincidence window with femtosecond temporal resolution needed to observe a coherent superposition of entangled photons with different energies. To take account of the detection crystals acceptance bandwidth, we define the modified joint spectral amplitude
\begin{equation}
\Gamma(\omega)\propto\Lambda(\omega)\Phi(\omega)
\end{equation}
with 
\begin{equation}\
\Phi(\omega)=\mathrm{sinc}\left[\frac{\left(\Delta k_{SFG}(\omega)-\frac{2\pi}{G_{SFG}}\right)L_{SFG}}{2}\right]
\end{equation}
and $\Delta k_{SFG}(\omega)$ now being the phase mismatch responsible for the efficiency of the SFG process. The recombined 532 nm photons are then imaged onto the photosensitive area of a SPCM (IDQ id100-50). In order to exclude the detection of residual infrared photons we mounted a bandpass filter in front of lens L4. The SFG signal $S$ is in general proportional to
\begin{equation}\label{eq:signal}
S\propto\left|\int_{-\infty}^{\infty}d\omega M(\omega)\Gamma(\omega)\right|^2.
\end{equation}
However, diffraction effects due to the finite aperture of the lenses and prisms lead to a point-to-spot image from $\Sigma_0$ to $\Sigma_1$. Therefore, a given frequency component is blurred over several pixels of the SLM. This effect is taken into account by a convolution of $\Gamma(\omega)$ with a Gaussian-shaped point spread function $PSF(\omega)$, i.e.,
\begin{equation}\label{eq:gamma_PSF}
\Gamma(\omega)\rightarrow\Gamma_{PSF}(\omega)\propto(\Gamma\otimes
PSF)(\omega)\ . 
\end{equation}  

\begin{figure}[ht]
\begin{center}
\fbox{\includegraphics[width=0.5\textwidth]{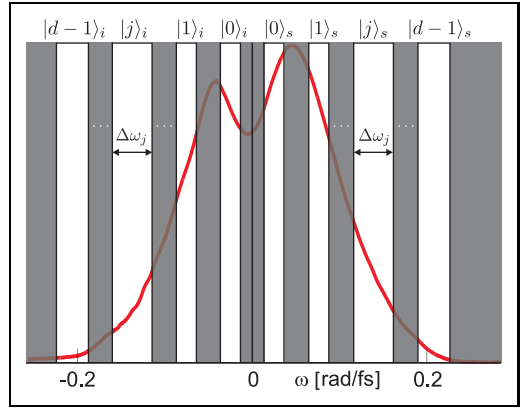}} 
\caption{\label{fig:freq_bins}Measured SPDC spectrum overlaid with a schematic frequency-bin pattern. The transmitted amplitude $\vert u_j^{i,s}\vert$ (white bars) of each bin can be adjusted through amplitude modulation by means of the SLM.}  
\end{center}
\end{figure}

\begin{figure}
  \centering
  \fbox{
    \includegraphics[width=0.75\linewidth]{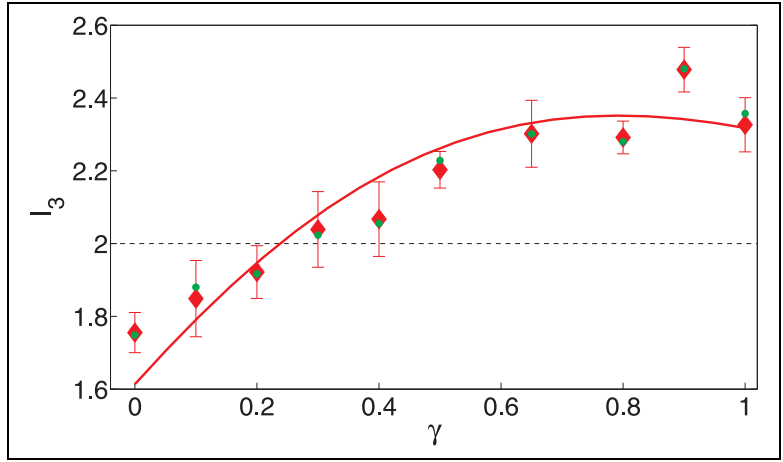}
  }
  \caption{Bell parameter $I_3(\gamma)$. The experimental Bell parameter $I^{\text{expt.}}_3$ is depicted with red diamonds. The $2\sigma$ uncertainties are calculated assuming Poisson statistics on background-subtracted coincidence counts. The theoretically predicted Bell parameter $I_3(\gamma)$ (solid red line) is scaled with its corresponding mixing parameter. The (horizontal) dashed black line indicates the local variable realistic. We experimentally determine the mixing parameter to be $ \lambda^{\text{expt.}}_3=0.807\pm0.008$ with  2$\sigma$ uncertainties. The green dots represent $I^{\text{expt.}}_3$ with signaling revised experimental data (discussion Section~\ref{vierzwei}).}
  \label{I3}
\end{figure}

We encode qudits in the frequency domain by projecting the continuous state state $\vert\psi\rangle$ of Eq.~\eqref{eq:DC_state} into a discrete $d^2$-dimensional subspace spanned by frequency-bin states $\vert j \rangle_i \vert k\rangle_s$ with $\left\vert j\right\rangle_{i,s} \equiv\int_{-\infty}^{\infty}d\omega f^{i,s}_j\left(\omega\right)\hat{a}^{\dagger}_{i,s}(\omega)\left\vert 0\right\rangle_{i,s}$ and $j=0,\ldots,d-1$. The frequency bins itself are defined according to
\begin{equation}\label{eq:fbins}
f_{j}^{i,s}(\omega) = \begin{cases} 1/\sqrt{\Delta\omega_j} &
  \text{for}\left|\omega-\omega_{j}\right|<\Delta\omega_j/2\ , \\
0 & \text{otherwise,} \end{cases} 
\end{equation}
and are represented in Figure \ref{fig:freq_bins} together with a measured SPDC spectrum.
Furthermore, imposing of the constraint $|\omega_j-\omega_k|>(\Delta\omega_j+\Delta\omega_k)/2$ for all $j,k$ ensures that adjacent bins do not overlap. Under the restriction of a monochromatic pump field, the projected state is expressed as 
\begin{equation}\label{eq:psidisc_tomo}
\vert\psi\rangle^{(d)} = \sum_{j=0}^{d-1} c_{j}\vert j \rangle_i \vert j\rangle_s
\end{equation}
with amplitudes $c_{j}=\int_{-\infty}^{\infty}d\omega f_{j}^{i}\left(\omega\right)f_{j}^{s}\left(-\omega\right)\Gamma(\omega)$. The frequency-bin structure of Eq.~(\ref{eq:fbins}) is applied on the SLM by the transfer function 
\begin{equation}\label{eq:mslm_tomo}
M^{i,s}(\omega)=\sum_{j=0}^{d-1}u_j^{i,s}f_{j}^{i,s}(\omega)=\sum_{j=0}^{d-1}|u_j^{i,s}|e^{i\phi_j^{i,s}}f_{j}^{i,s}(\omega)\
,
\end{equation}
where the amplitude $|u_j^{i,s}|$ and phase $\phi_j^{i,s}$ of bin $j$ is controlled independently. Since in our experiment there is no spatial separation between idler and signal modes, we address each photon individually by assigning $M^i(\omega)$ to the lower-frequency part and $M^s(\omega)$ to the higher-frequency part of the spectrum. In fact, due to the PSF, we do have a small overlap between the spectral components of the idler and the signal domain such that the two spaces cannot be considered completely independent of one another. The measured signal of Eq.~\eqref{eq:signal} is equivalent to the projection
\begin{equation}\label{eq:projection_tomo}
S=\left\vert\langle\chi\vert\psi\rangle^{(d)}\right\vert^2=\left\vert\sum_{j=0}^{d-1}u^i_j u^s_j c_j\right\vert^2 
\end{equation}
for a direct product state 
\begin{equation}\label{eq:chi}
\vert\chi\rangle=\left(\sum_{j=0}^{d-1}u^{i*}_j\vert j\rangle_i\right)\left(\sum_{j'=0}^{d-1}u^{s*}_{j'}\vert j'\rangle_s\right)
\end{equation}
and a decomposition of the transfer function according to
Eq.~\eqref{eq:mslm_tomo}. The combination of the SLM together with an
SFG coincidence detection therefore realizes a projective
measurement. Different quantum-information protocols can thus be
implemented through a judicious choice of $\vert\chi\rangle$ together
with Eq.~(\ref{eq:mslm_tomo}). The state of Eq.~\eqref{eq:chi} can,
 for instance,
 be chosen in the form of a tomographically complete set to perform quantum state reconstruction of maximally entangled qudits up to $d=4$ as demonstrated in \cite{Bernhard2013}. Here, Eq.\eqref{eq:chi} serves to perform Bell measurements for maximally and non-maximally entangled qutrits. If we identify $i\leftrightarrow A$ and $s\leftrightarrow B$, the Bell parameter of Eq.~\eqref{eq:I3} is described by a combination of projective measurements onto the state of Eq.~\eqref{eq:chi} expressed as $\vert\chi_{a,b}\rangle=\vert a\rangle_A^{x}\vert b\rangle_B^{y}$, and defined by
\begin{equation}\label{eq:basisA}
\vert
a\rangle_A^{x}=\frac{1}{\sqrt{3}}\sum_{j=0}^{2}\exp\left(i\frac{2\pi}{3}j(a+\alpha_x)\right)\vert
j\rangle_A\ ,
\end{equation}
\begin{equation}\label{eq:basisB}
\vert
b\rangle_B^{y}=\frac{1}{\sqrt{3}}\sum_{j=0}^{2}\exp\left(i\frac{2\pi}{3}j(-b+\beta_y)\right)\vert
j\rangle_B\ ,
\end{equation}
with $a,b=0,1,2$ for a specific choice of detection settings
$\alpha_1=0, \alpha_2=1/2, \beta_1=1/4,$ and $\beta_2=-1/4$ according
to \cite{Collins2002}. Note, that these settings are only optimal in
the case of $\gamma=1$ and $\gamma=\gamma_{max}$~\cite{Acin2002}. Allowing for the
state of Eq.~(\ref{eq:qutrit_gamma}), the individual joint probabilities become a function of $\gamma$. In accordance with Eq.~(\ref{eq:projection_tomo}), a measured coincidence signal is then given by 
\begin{equation}\label{system}
P_{\gamma}(A_x=a,B_y=b)\propto\left\vert\langle\chi_{a,b}\vert\psi(\gamma)\rangle^{(3)}\right\vert^2 
\end{equation}
with the projection states of Eq.~\eqref{eq:basisA} and \eqref{eq:basisB}. Starting from a maximally entangled state through Procrustean filtering \cite{Bennett1996}, the reduced entanglement, i.e.,~$\gamma<1$, is obtained by decreasing the transmission amplitudes $\vert u_1^{i}\vert$ and $\vert u_1^{s}\vert$ of the bins associated with  $\vert 1 \rangle_A \vert 1\rangle_B$ using the SLM. The experimental data $I^{\text{expt.}}_3(\gamma)$ (red diamonds) are depicted in Figure~\ref{I3} together with the theoretical result for $I_3(\gamma)$ (solid red line). The latter is scaled to the measured data using a symmetric noise model   
$\hat{\rho}^{(d)}=\lambda_d\vert\psi\rangle^{(d)}\,^{(d)}\langle\psi\vert+(1-\lambda_d)\mathbbm{1}_{d^2}/{d^2}$,
where the mixing parameter $\lambda_d$ quantifies the deviations from a pure state and $\mathbbm{1}_{d^2}$ denotes a $d^2$-dimensional identity operator. Under the given measurement settings together with white noise and possible misalignments in the experimental setup Figure \ref{I3} demonstrates a violation of Eq.~\eqref{eq:CGLMP} for $\gamma \geq 0.5$. 

\section{Measures of Non-Locality}

\subsection{Systems and correlations}

The goal of the experiment discussed in this article is to understand and quantify {\em correlations\/} that
occur when two parts of a physical system in a common state are measured. More precisely, this type of correlation
displays itself in the behavior of a {\em system}, i.e., a joint
input-output behavior. 
%In this paper we will use different methods for analyzing quantum
%behaviors. In some of our analysis we are only interested in the
%probabilities we get from quantum physics and, therefore, we need
%some definitions to talk about them.
\begin{defi}
{\rm 
A two-party {\em input-output behavior\/} or  {\em system\/} is a
conditional probability distribution $P(ab|xy)$, where $x$ and $y$
are the respective inputs
of the two parties and $a$ and $b$ are their outputs.
}\end{defi} 
A {\em two-partite quantum state\/} is a natural way to obtain a 
  system: Given a set of possible alternative measurements ($x$ and
$y$) to be
carried out by the two parties, their outcomes ($a$ and $b$)
correspond to  the outputs of the system. We establish a general
framework for studying such systems, and for quantifying their non-locality,
independently of whether they come from a quantum state or not.

Classically speaking, the two players not allowed to communicate can, for obtaining a given 
joint behavior, apply 
{\em local strategies}. Generally, such a strategy can employ local randomness
and even shared (classical) information. 
\begin{defi}
{\rm 
Two players' {\em local strategies\/} are ways of
determining their respective outputs as a function of  their inputs.
A {\em local strategy (for the first player)\/} is of the form
$a=f(x,L,R)$, where $f$ is a function, $L$ is a classical random
variable and $R$ is a shared (i.e., also known to the other player)
random variable. 
As a special case, a strategy is {\em local deterministic\/} if it is 
of the form $a=f(x)$ for some fixed function $f$. 
}\end{defi} 
A system that cannot be simulated by any pair of local strategies is
called non-local.
Such systems are, in classical terms, only explainable 
by {\em communication}. It is important to note, however, that 
this does not mean that any non-local system {\em allows\/}
for such communication. Actually, our focus is precisely on the systems 
which are non-local yet non-signaling at the same time. 
\begin{defi}
{\rm 
A two-partite system is {\em local\/} if  it can be simulated by two
players using a pair of local strategies.
It is {\em non-signaling\/} if the output of a party, given its
input, is independent of the input of the other party: We have
$P(a|xy)=P(a|x)$, and {\em vice versa}.
}\end{defi}

%This behavior is what we call a \textit {input-output behavior} $G$ with input $I$ and output $O$. Input $I$ and output $O$ can be considered tuples $I=\{I_1,I_2,....,I_n\}$, $O=\{ O_1,O_2,...O_n\}$, where $n$ is the number of parties. 
%And the whole behavior can be described by a probability distribution $P(O_1,O_2,...,O_n | I_1,I_2,...,I_n)$.

 %Signaling means that a given behavior could be used to send signals from one party to another (by choosing the input accordingly). The different parties are often spatially separated, and since we assume faster than light communication to be impossible, we want that our behaviors are \textit{non-signaling}. Non-signaling means that no party can (by choosing its input) send messages to other parties.
%Mathematically spoken this means that $\forall_{j}, \forall_{k \neq j} P(O_j|I_k)=P(O_j).$

%A input output behavior between two or more spatially separated parties can be considered \textit{local}, when it can be simulated by the parties without communication using only classical shared random variables. The parties can, therefore, only use \textit{local} strategies, i.e. there output can only depend on the classical shared random variable and its own input.

It is the goal of the rest of this section to identify criteria for
testing whether a (non-signaling) system is local or not and, if not,
to give a measure for the strength or amount of non-locality it
displays. 
Whereas the violation of a fixed Bell inequality is a sufficient
criterion for non-locality, it is {\em not necessary}. In the same sense, 
the extent by which a given inequality is violated does not 
directly measure non-locality. We describe two criteria and 
measures: The first corresponds, in a sense, to taking {\em all\/} Bell
inequalities into account, whereas the second is of different nature
and  based on the price, measured in terms of classical communication, one
has to pay for establishing a correlation.

\subsection{Distance to the polytope of local correlations}

Within the space of non-signaling two-party systems, the {\em local\/}
systems form a subset or, more precisely, a {\em convex
  polytope}, somewhat abusively called the {\em local polytope}. Whereas
Bell inequalities characterize this polytope in terms of its {\em
  facets}, we can also do the same through its extremal points, which
correspond to the {\em local deterministic strategies\/}; every 
local strategy is a convex combination of local deterministic
strategies. We use this description as a locality test as well as a
quantification 
of the non-local content of a given, e.g., experimentally  observed, system.

For a system $P(ab|xy)$, we denote by $|a|$, $|b|$, $|x|$, and $|y|$
the sizes of the ranges of the corresponding random variables.
A pair of strategies can be represented in the $(|a||x||b||y|)$-dimensional 
real space. The number of pairs of local deterministic strategies is 
$|a|^{|x|}|b|^{|y|}$. 
Now, any local (probabilistic) strategy is  a (non-trivial) convex
combination of local deterministic strategies. We check
whether an observed  behavior lies inside this polytope, and if not,
we calculate the distance to this polytope in the $L^1$ norm, using {\em linear programming} \cite{dantzig1965,elliott2009}, by 
minimizing the difference between the given behavior and a local
approximation of it.
The larger this distance, the stronger a behavior can be considered
non-local. 
More specifically, the fact that a behavior has non-zero
distance to the polytope can indicate that it violates a Bell
inequality 
{\em or\/} that it is signaling (or both).

%Author: Marcel
%\begin{itemize}
%\item General idea / what means local.
%\item What is the local polytope?
%\item How to build it from the local strategies?
%\item How to calculate the distance?
%\end{itemize}

\subsection{The amount of communication required for establishing correlations}
%Author: Alberto
%\begin{itemize}
%\item What do we mean with communication complexity?
%\item What does the communication complexity mean?
%\item Connection to simulation of a quantum channel.
%\item Why is a non zero value interesting?

%\end{itemize}

A non-locality measure alternative to Bell-inequality or
polytope-distance based measures
uses the fact that the local correlations are exactly the ones that
can,
classically speaking, be explained by shared information and  
require {\em no\/} communication. Hence, a non-locality measure 
 is given by the {\em minimal amount\/} of classical 
communication required to simulate the correlations. In \cite{mw1}, 
 this measure has been called {\it non-local capacity}. There, it was proven that the non-local
capacity is the minimum of a convex functional over a suitable space of
probability distributions, so that its computation is a convex-optimization
problem. We  review that method, which we apply for
calculating the non-local capacity from the experimental data.

\subsubsection{Communication cost and non-local capacity.}

A general 
classical simulation of correlations $P(ab|xy)$ employing a one-way communication
between the two parties is as follows. A party, say Alice, chooses the input $x$
corresponding to the measurement she wants to simulate. She generates a 
variable $k$ and the measurement outcome $a$ according to a conditional probability 
distribution $P(ak|xr)$ depending on the input~$x$ and a random variable 
$R$ shared with the other party, say Bob, and generated with probability
distribution $\rho(R)$. Then, she sends $k$ to Bob. Finally, 
Bob chooses the measurement he wants to simulate, labelled by the index $y$, and
 generates an outcome~$B$ according to a conditional probability
distribution $\rho(b|ykr)$. The protocol exactly simulates the correlations
$P(ab|xy)$ if
\be
\begin{array}{c}
\sum_k\int dr P(b|ykr)P(ak|xr)\rho(r)=P(ab|xy)\ .
\end{array}
\ee

There are different definitions of communication cost of a simulation. We could
define the communication cost as the number of required bits in the worst
case, or as the average number of bits. 
As done in \cite{mw1}, we employ the entropic definition and we
define the communication cost, say $\cal C$, as the maximum, over the 
space of distributions $P(a)$, of the conditional Shannon entropy
$H(K|Y)\equiv-\int dr \rho(r)\sum_k P(k|r)\log_2 P(k|r)$.
That is,
\be\label{comm_cost}
{\cal C}\equiv\max_{P(x)} H(K|R)\ .
\ee 
We define the {\it non-local capacity}, denoted by ${\cal C}_{nl}$, of the
correlations $P(ab|xy)$ as the minimal amount of communication $\cal C$ 
required for an exact simulation of them.
More generally, we can perform a parallel simulation of $N$ distinct pairs of 
entangled systems. We define the {\em asymptotic communication cost}, denoted 
by ${\cal C}^{asym}$, as the communication cost of the simulation divided by $N$, 
in the limit $N\rightarrow\infty$ (see \cite{mw1} for a more detailed
definition). The asymptotic non-local capacity, denoted by ${\cal C}_{nl}^{asym}$, 
is defined as the minimum of ${\cal C}^{asym}$ over the class of parallel
simulations.

\subsubsection{A convex-optimization problem.}

In the following, we  assume that Bob can choose a measurement among
a finite set of possibilities. The index $y$ can take values from $1$
to $M$.
In \cite{mw1}, it was shown that the asymptotic non-local capacity is the
minimum of a convex functional over a suitable space of probability distributions
$\cal V$. 
\begin{defi}
{\rm 
Given a non-signaling system $P(ab|xy)$,
the set ${\cal V}$ contains any conditional probability $\rho(a{\bf b}|x)$ 
over $a$ and the sequence ${\bf b}=(b_1,\dots,b_M)$ whose marginal distribution of 
$a$ and the $m$-th variable is the distribution $P(ab|x,y=m)$.
}
\end{defi}

In other words, ${\cal V}$ contains any $\rho(a{\bf b}|x)$ satisfying the 
constraints
\be\label{constraints}
\sum_{{\bf b},b_m=b} \rho(x{\bf b}|x)=P(ab|x,y=m)\ ,
\ee
where
\be
\sum_{{\bf b},b_m=b}   \rightarrow  \sum_{b_1,\dots,b_m=b,\dots,b_M}
\ee
is the summation over every index in $\bf b$ but the $m$-th one, which is set equal to 
$b$.

We proved that
\be
{\cal C}_{nl}^{asym}=\min_{\rho(a{\bf b}|x)\in{\cal V}} {\cal
  C}(x\rightarrow{\bf b})\ ,
\ee
where 
\be
{\cal C}(a\rightarrow{\bf b})\equiv \max_{\rho(x)} I({\bf b};x)
\ee
is the capacity of the channel $\rho({\bf b}|x)$. Let us recall that the capacity of
a channel $a\rightarrow {\bf b}$ is the maximum of the mutual information $I({\bf b};x)$
between the input and the output over the space of input probability distributions
$\rho(x)$~\cite{cover}.
As the mutual information is convex and the maximum over a set of convex functions
is still convex~\cite{boyd}, the asymptotic communication complexity is the
minimum of a convex function over the space $\cal V$. Furthermore, the equality constraints
defining the set $\cal V$ are linear. This implies that the minimization problem is convex
and can be numerically solved with standard methods~\cite{boyd}.

\section{Analysis of Experimental Data}
\subsection{The raw data display signaling}

\begin{figure}[]
  \centering
  \fbox{
    \includegraphics[width=1\linewidth]{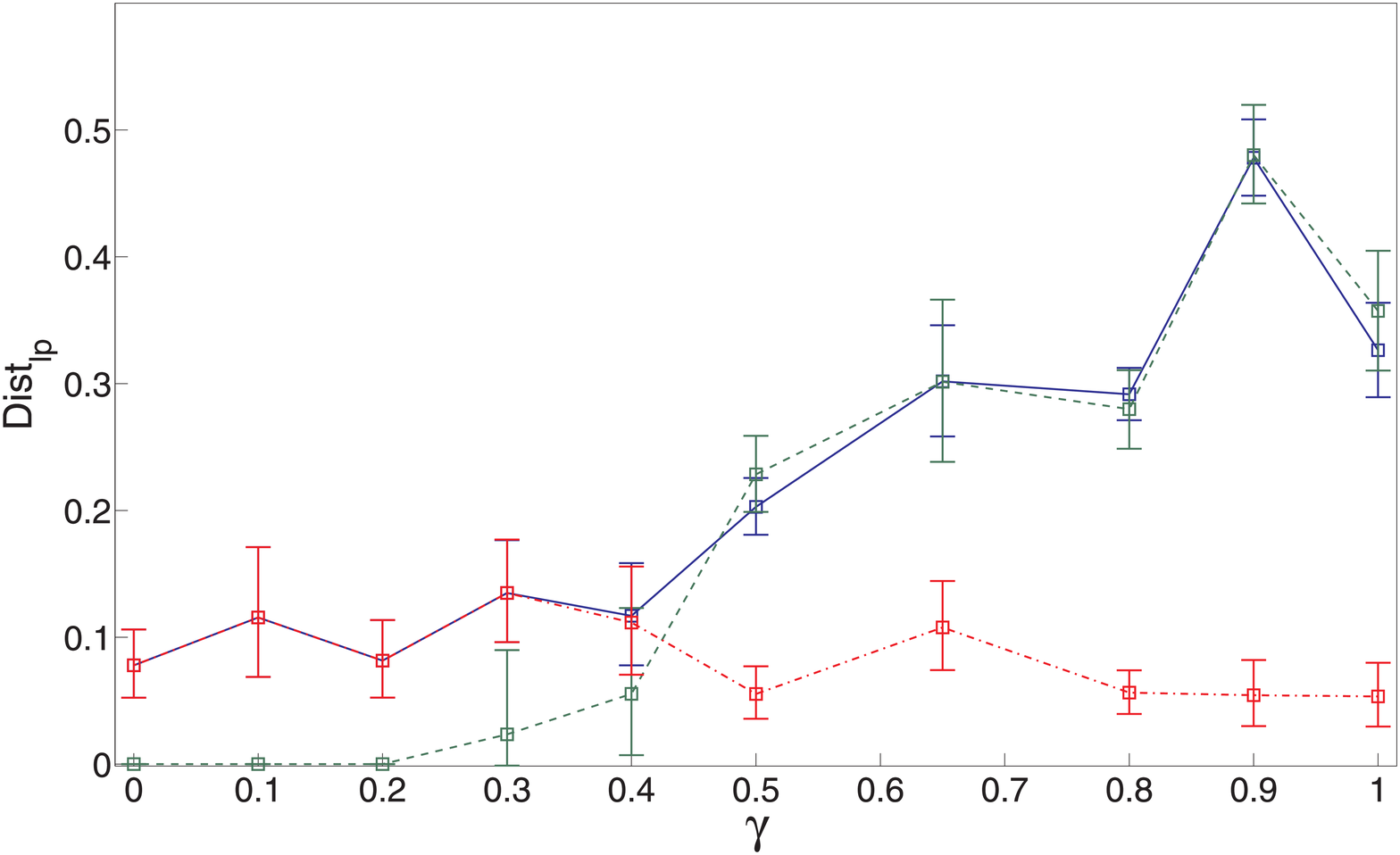}
  }
  \caption{The distance to the local polytope ($Dist_{lp}$) before (blue squares) and
after (green squares) removing the signaling part, and 
the
distance to the non-signaling polytope (red squares). The corresponding lines (solid blue, dashed green, dot-dashed red) are guides to the eyes.}
  \label{FG1}
\end{figure}

We analyze the experimentally obtained two-partite system
according to Eq.~(\ref{system}). 
A~first remarkable result of our
analysis 
of this system is that it happens to be, actually, not only non-local but even
{\em signaling}. From the point of view of how the qutrits were
experimentally realized (not spatially separated), this is not overly 
surprising and, in particular, not in violation of relativity.
In order to be able to apply the non-locality quantification, designed for 
non-signaling systems,  to the data, we free them of their
signaling part, replacing them by the closest respective
non-signaling systems. Figure \ref{FG1} demonstrates that for
$\gamma<0.2$, the 
resulting non-signaling systems end up being local, whereas for
$\gamma\geq0.5$, there is no significant difference between the raw and
the corrected data.

%But what is interesting is, that when we calculate the distance to the local polytope from the original data, we are non-local for the whole range of $\gamma$,
%which is not the case, when we consider the Bell inequality violation.
%Also it is quite interesting to see, that for bigger $\gamma$s the signaling not only has a smaller effect on the whole behavior, but is also smaller in general.
%For us this effect is quite desirable, since this justifies removing signaling for big $\gamma$s and using the data without signaling for further analysis, 
%including the calculation of the communication complexity.

\subsection{Removal of the signaling part}\label{vierzwei}

Removing the signaling part in the described way can be seen as some kind of error
correction; after all, we know that quantum theory is, despite the
phenomena of entanglement and non-locality, non-signaling. 
The computation of the non-signaling part is performed similarly to
the computation of the distance to the polytope, where in this case,
the {\em non-signaling polytope\/} instead of the local polytope is used.
Our method of getting rid of experimental errors is similar to a
procedure discussed in the literature~\cite{ChristandlRenner} that
proposes
a maximum-likelihood estimation from the set of all quantum states.
We adapt the method for non-signaling systems instead of physical
states. 
Note that while the removal of the signaling part can greatly
influence the measures for non-locality used here, it has almost 
no influence on Bell-inequality violations (Figure~\ref{I3}).

%Before analyzing the non-signaling data we want to make another comparison between this data and the original data. 
%For this purpose we compared the Bell inequality violation with and without signaling and we can make two interesting observations.
%First removing the signaling part has almost no influence on the Bell value and second also here the effect is smaller when the $\gamma$ is bigger.
%For the big $\gamma$s the effect is even almost negligible.

\subsection{Comparison between the two methods}

\begin{figure}[]
  \centering
  \fbox{
    \includegraphics[width=1\linewidth]{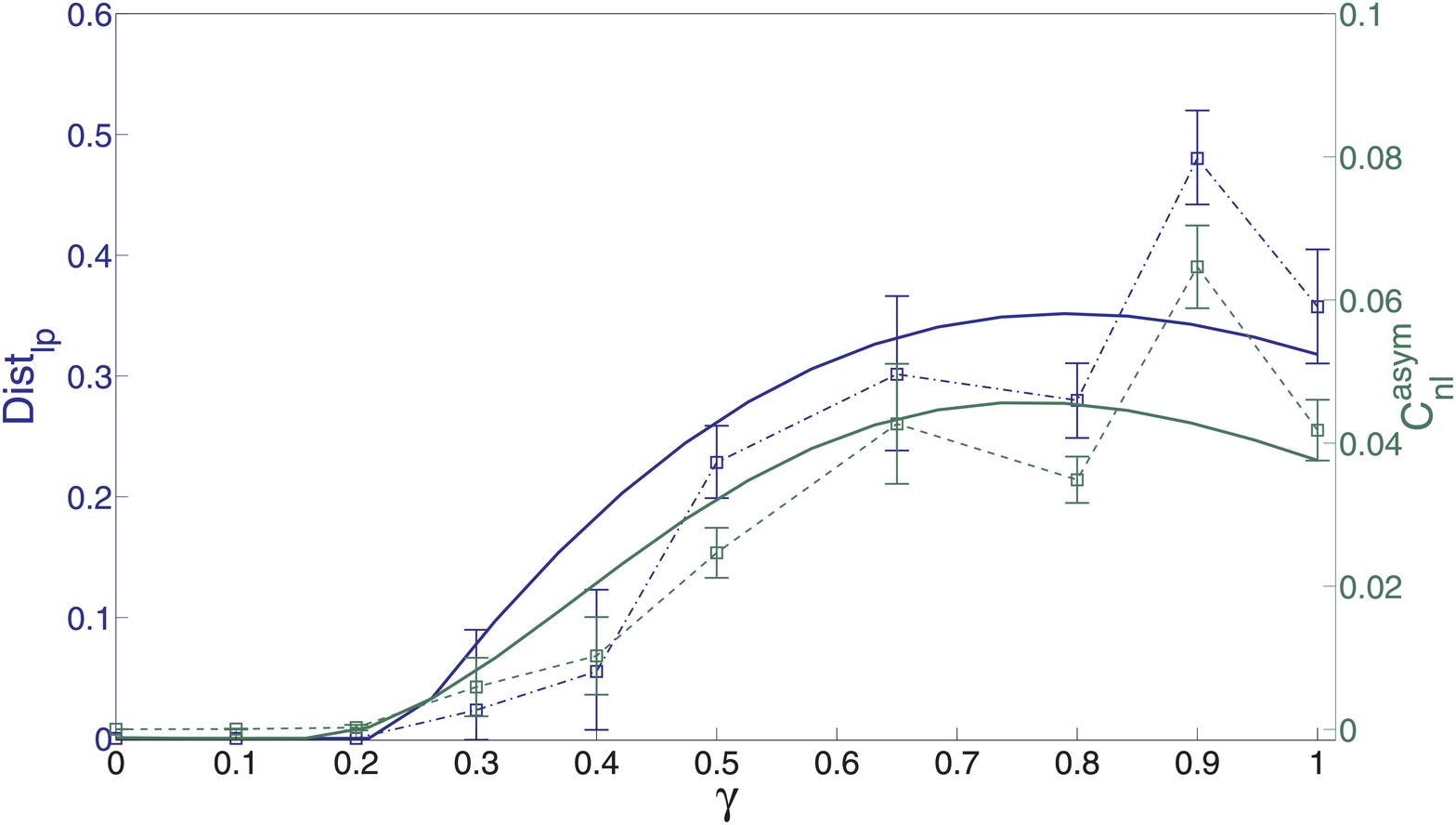}
  }
  \caption{Two alternative methods for quantifying non-locality are compared to each other. The blue squares with the blue scale on the left show the distance to the local polytope ($Dist_{lp}$), whereas the green squares with the green scale on the right show the asymptotic non-local capacity ($C^{asym}_{nl}$). The corresponding lines (dot-dashed blue, dashed green) are guides to the eyes. Theoretical values for $Dist_{lp}$ and $C^{asym}_{nl}$ are indicated with solid lines (upper blue, lower green).}
  \label{FG2}
\end{figure}

First, we observe that the two non-locality measures behave very
similarly for the investigated measurement data (Figure \ref{FG2}). Second, we can
conclude 
from our analysis that the {\em non-monotonicity\/} of the measured
non-locality 
as compared to the {\em entanglement\/} | a phenomenon already
reported previously in the context of non-maximally entangled qubit
pairs~\cite{BrunnerGisin} | is not merely an effect that is respective
to a single (strange) Bell inequality, but that persists also with
respect to the non-locality measures.

\section{Concluding Remarks}
%Author: Stefan
%\begin{itemize}
%\item Implementation of qutrit
%\item Could show the special properties of the Bell equality in the experiment.
%\item A new reliable simple and fast method for non-locality test. (Shows similar method than polytope method.)
%\item Bell inequalities don't show everything.
%\item Signaling in experimental data can exist (statistical error but also others). One has to deal with this.
%\item Still dependent of the measurement bases (and Bell inequalities can help finding the best bases).
%\item Such tools could help to improve measurement-settings (analyzing non-locality and signaling).
%\item We could show a for a certain types of states and some fixed measurements, that the non-locality does not behave monotonous in respect to the entanglement.
%This is true even if we look at the distance to the non-local polytope and the communication cost, which is a stronger statement then only looking at a certain Bell inequality! 
%\end{itemize}
In this article, dedicated to the celebration of the 50th
anniversary of John Stewart Bell's famous and seminal theorem, we have
described an experiment for measuring the non-locality of
entangled qutrit pairs. We have analyzed the resulting experimental data,
and in particular the strength of their non-locality, with two non-standard methods,
namely the distance to the local polytope (in the space of non-signaling behaviors)
as well as the communication cost for classically simulating the correlations.

In the results presented here, there is no difference between the
non-local content 
as measured by the Bell parameter compared to the alternative methods, i.e., the distance to the local polytope and the non-local capacity.
This is because the measurement settings are chosen in order to
strongly  violate 
a specific Bell
inequality. 
However, for an arbitrary state for which
the optimal 
Bell inequality is not known, the alternative methods offer a 
more effective way of testing non-locality, by performing less measurements.

The phenomenon of non-locality, John Bell's profound discovery,
continues to put into question the way we
traditionally
view space and time. 
When one tries to understand --- just as Bell himself did in a number
of articles~\cite{BellSU} --- non-locality in the context of different interpretations of quantum theory,
the conclusion is always the same: It does not fit. In collapse theories, how can spontaneously generated
information be identical in different spatial positions? In modern deterministic theories such as
the ``church of the larger Hilbert space,'' parallel universes or ``parallel lives''~\cite{BraRob}, no mechanism
that could {\em explain\/} the correlations has been described. It appears
that {\em the\/} interpretation which passes the ``Bell test'' and is suitable  for embedding, in a
non-artificial way, non-local correlations is yet to be found.
It may have the property that space and time, in particular spatial
distance and spacelike separation, do not exist prior to, but only 
emerge {\em together\/} with the (correlated) classical information.
\\ \ \\
%{\bf Acknowledgements.} This research was supported by the Swiss
%National Science Foundation (SNF), the NCCR {\em QSIT}, the COST
%action on ``Fundamental Problems in Quantum Physics,'' and the
%CHIST-ERA {\em DIQIP} as well as the Swiss National Science Foundation
%Grant No.~PP00P2\_133596 and the NCCR {\em MUST}. 
{\bf Acknowledgements.} This research was supported by
 the Swiss
National Science Foundation (SNF), the NCCR {\em QSIT}, the NCCR {\em MUST}, the SNF Grant No.~PP00P2\_133596,
 the COST
action on ``Fundamental Problems in Quantum Physics,'' and the
CHIST-ERA {\em DIQIP}. 
\\
\
\\

\bibliographystyle{plain} 
\bibliography{literature}

\begin{thebibliography}{10}

\bibitem{Acin2002}
Antonio Ac\'in, Thomas Durt, Nicolas Gisin, and Jos\'e~I Latorre.
\newblock Quantum nonlocality in two three-level systems.
\newblock {\em Physical Review A}, 65:052325, 2002.

\bibitem{Aspect1982a}
Alain Aspect, Jean Dalibard, and G{\'e}rard Roger.
\newblock Experimental test of {B}ell's inequalities using time-varying
  analyzers.
\newblock {\em Physical Review Letters}, 49(25):1804--1807, 1982.

\bibitem{transitiv}
Tomer~J Barnea, Jean-Daniel Bancal, Yeong-Cherng Liang, and Nicolas Gisin.
\newblock A tripartite quantum state violating the hidden influence
  constraints.
\newblock {\em Physical Review A}, 88:022123, 2013.

\bibitem{Bell1964}
John~S Bell.
\newblock {On the {E}instein-{P}odolsky-{R}osen paradox}.
\newblock {\em Physics}, 1:195--200, 1964.

\bibitem{BellSU}
John~S Bell.
\newblock {\em Speakable and Unspeakable in Quantum Mechanics: Collected Papers
  on Quantum Philosophy}.
\newblock Cambridge University Press, 2004.

\bibitem{Bennett1996}
Charles~H Bennett, Herbert~J Bernstein, Sandu Popescu, and Benjamin Schumacher.
\newblock Concentrating partial entanglement by local operations.
\newblock {\em Physical Review A}, 53(4):2046, 1996.

\bibitem{Bernhard2013}
Christof Bernhard, B\"anz Bessire, Thomas Feurer, and Andr\'e Stefanov.
\newblock Shaping frequency-entangled qudits.
\newblock {\em Physical Review A}, 88:032322, 2013.

\bibitem{boyd}
Stephen Boyd and Lieven Vandenberghe.
\newblock {\em Convex Optimization}.
\newblock Cambridge University Press, 2004.

\bibitem{BraRob}
Gilles Brassard and Paul Raymond-Robichaud.
\newblock Can free will emerge from determinism in quantum theory?
\newblock In {\em Is Science Compatible with Free Will?}, pages 41--61.
  Springer, 2013.

\bibitem{BrunnerGisin}
Nicolas Brunner, Nicolas Gisin, and Valerio Scarani.
\newblock Entanglement and non-locality are different resources.
\newblock {\em New Journal of Physics}, 7(1):88, 2005.

\bibitem{ChristandlRenner}
Matthias Christandl and Renato Renner.
\newblock Reliable quantum state tomography.
\newblock {\em Physical Review Letters}, 109(12):120403, 2012.

\bibitem{Clauser1969}
John~F Clauser, Michael~A Horne, Abner Shimony, and Richard~A Holt.
\newblock Proposed experiment to test local hidden-variable theories.
\newblock {\em Physical Review Letters}, 23(15):880--884, 1969.

\bibitem{Collins2002}
Daniel Collins, Nicolas Gisin, Noah Linden, Serge Massar, and Sandu Popescu.
\newblock {B}ell inequalities for arbitrarily high dimensional systems.
\newblock {\em Physical Review Letters}, 88(4):040404, 2002.

\bibitem{cover}
Thomas~M Cover and Joy~A Thomas.
\newblock {\em Elements of Information Theory (Wiley Series in
  Telecommunications and Signal Processing)}.
\newblock Wiley-Interscience, 2006.

\bibitem{dantzig1965}
George~B Dantzig.
\newblock {\em Linear programming and extensions}.
\newblock Princeton university press, 1965.

\bibitem{dayan_silberberg05prl}
Barak Dayan, Avi PeÕer, Asher~A Friesem, and Yaron Silberberg.
\newblock Nonlinear interactions with an ultrahigh flux of broadband entangled
  photons.
\newblock {\em Physical Review Letters}, 94(4):043602, 2005.

\bibitem{elliott2009}
Matthew~B Elliott.
\newblock A linear program for testing local realism.
\newblock {\em arXiv preprint arXiv:0905.2950}, 2009.

\bibitem{Freedman1972}
Stuart~J Freedman and John~F Clauser.
\newblock Experimental test of local hidden-variable theories.
\newblock {\em Physical Review Letters}, 28(14):938--941, 1972.

\bibitem{Giustina2013}
Marissa Giustina et~al.
\newblock {B}ell violation using entangled photons without the fair-sampling
  assumption.
\newblock {\em Nature}, 497(7448):227--230, 2013.

\bibitem{Hofmann2012}
Julian Hofmann, Michael Krug, Norbert Ortegel, Lea G{\'e}rard, Markus Weber,
  Wenjamin Rosenfeld, and Harald Weinfurter.
\newblock Heralded entanglement between widely separated atoms.
\newblock {\em Science}, 337(6090):72--75, 2012.

\bibitem{Horodecki1995}
Ryszard Horodecki, Pawe{\l} Horodecki, and Micha{\l} Horodecki.
\newblock Violating {B}ell inequality by mixed spin-$\frac{1}{2}$ states:
  Necessary and sufficient condition.
\newblock {\em Physics Letters A}, 200(5):340--344, 1995.

\bibitem{KS}
Simon Kochen and Ernst Specker.
\newblock The problem of hidden variables in quantum mechanics.
\newblock In {\em The Logico-Algebraic Approach to Quantum Mechanics}, pages
  293--328. Springer, 1975.

\bibitem{mw1}
Alberto Montina and Stefan Wolf.
\newblock Information-based measure of nonlocality.
\newblock {\em arXiv preprint arXiv:1312.6290}, 2013.

\bibitem{RenWol}
Renato Renner and Stefan Wolf.
\newblock Quantum pseudo-telepathy and the {K}ochen-{S}pecker theorem.
\newblock In {\em IEEE International Symposium on Information Theory}, pages
  322--322, 2004.

\bibitem{Rowe2001a}
Mary~A Rowe, David Kielpinski, V~Meyer, Charles~A Sackett, Wayne~M Itano,
  Christopher Monroe, and David~J Wineland.
\newblock Experimental violation of a {B}ell's inequality with efficient
  detection.
\newblock {\em Nature}, 409(6822):791--794, 2001.

\bibitem{Salart2008}
Daniel Salart, Augustin Baas, Cyril Branciard, Nicolas Gisin, and Hugo Zbinden.
\newblock Testing the speed of Ôspooky action at a distanceÕ.
\newblock {\em Nature}, 454(7206):861--864, 2008.

\bibitem{Stefanov2002}
Andr{\'e} Stefanov, Hugo Zbinden, Nicolas Gisin, and Antoine Suarez.
\newblock Quantum correlations with spacelike separated beam splitters in
  motion: Experimental test of multisimultaneity.
\newblock {\em Physical Review Letters}, 88(12):120404, 2002.

\bibitem{Tittel1998}
Wolfgang Tittel, Juergen Brendel, Hugo Zbinden, and Nicolas Gisin.
\newblock Violation of {B}ell inequalities by photons more than 10 km apart.
\newblock {\em Physical Review Letters}, 81(17):3563, 1998.

\bibitem{Ursin2007}
Rupert Ursin et~al.
\newblock Entanglement-based quantum communication over 144 km.
\newblock {\em Nature Physics}, 3(7):481--486, 2007.

\bibitem{Weihs1998}
Gregor Weihs, Thomas Jennewein, Christoph Simon, Harald Weinfurter, and Anton
  Zeilinger.
\newblock Violation of {B}ell's inequality under strict {E}instein locality
  conditions.
\newblock {\em Physical Review Letters}, 81(23):5039, 1998.

\bibitem{Weiner2000}
Andrew~M Weiner.
\newblock {Femtosecond pulse shaping using spatial light modulators}.
\newblock {\em Review of Scientific Instruments}, 71(5):1929--1960, 2000.

\bibitem{Wihler2012}
Thomas~P Wihler, B{\"a}nz Bessire, and Andr{\'e} Stefanov.
\newblock Computing the entropy of a large matrix.
\newblock {\em arXiv preprint arXiv:1209.2575}, 2012.

\bibitem{Zohren2008}
Stefan Zohren and Richard~D Gill.
\newblock Maximal violation of the
  {C}ollins-{G}isin-{L}inden-{M}assar-{P}opescu inequality for infinite
  dimensional states.
\newblock {\em Physical Review Letters}, 100(12):120406, 2008.

\end{thebibliography}
%\bibliography{biblio.bib}
%\begin{thebibliography}{10}

%\bibitem{mw1} A. Montina, S. Wolf, arXiv:1312.6290.
%\bibitem{cover} T. M. Cover and J. A. Thomas, {\it Elements of Information Theory} (Wiley, New York, 1991).
%\bibitem{boyd} S. Boyd, L. Vandenberghe, {\it Convex Optimization} 
%(Cambridge University Press, Cambridge, 2004).

%\end{thebibliography}

\end{document}